\documentstyle[12pt,epsfig]{article}

\newcommand{\be}{\begin{equation}}
\newcommand{\ee}{\end{equation}}
\def\aprle{\buildrel < \over {_{\sim}}}
\def\aprge{\buildrel > \over {_{\sim}}}
\begin{document}
\topmargin 0pt
\oddsidemargin=-0.4truecm
\evensidemargin=-0.4truecm
\renewcommand{\thefootnote}{\fnsymbol{footnote}}
\newpage
\setcounter{page}{1}
\begin{titlepage}     
\vspace*{-2.0cm}
\begin{flushright}
FISIST/12-99/CFIF \\
hep-ph/9907435
\end{flushright}
\vspace*{0.5cm}
\begin{center}
\vspace*{0.2cm}
{\Large \bf Parametric resonance in neutrino oscillations in matter}
$\!$ \footnote{Invited review article to be published in the special 
issue of {\it Pramana} dedicated to neutrino physics} \\
\vspace{1.0cm}

{\large E. Kh. Akhmedov
\footnote{On leave from National Research Centre Kurchatov Institute, 
Moscow 123182, Russia. E-mail: akhmedov@gtae2.ist.utl.pt}}\\
\vspace{0.05cm}
{\em Centro de F\'\i sica das Interac\c c\~oes Fundamentais (CFIF)} \\
{\em Departamento de Fisica, Instituto Superior T\'ecnico}\\
{\em Av. Rovisco Pais, P-1049-001 Lisboa, Portugal}\\
\end{center}
\vglue 1.2truecm
\begin{abstract}
Neutrino oscillations in matter can exhibit a specific resonance 
enhancement -- parametric resonance, which is different from the 
MSW resonance. Oscillations of atmospheric and solar neutrinos 
inside the earth can undergo parametric enhancement when neutrino 
trajectories cross the core of the earth. In this paper we review  
the parametric resonance of neutrino oscillations in matter. 
In particular, physical interpretation of the effect and the prospects 
of its experimental observation in oscillations of solar and atmospheric
neutrinos in the earth are discussed. 
\end{abstract}
\end{titlepage}
\renewcommand{\thefootnote}{\arabic{footnote}}
\setcounter{footnote}{0}
\newpage
\section{Introduction}
It is well known that neutrino oscillations in matter can differ significantly  
from oscillations in vacuum, the best studied example being the 
Mikheyev-Smirnov-Wolfenstein (MSW) effect \cite{MS,W}. It is, however, much 
less known that the MSW effect is not the sole mechanism by which matter can 
enhance transitions between neutrinos of different flavor. The MSW effect 
enhances the probabilities of neutrino flavor transitions by amplifying 
neutrino {\it mixing}: the mixing angle in matter $\theta$ can become
equal to $45^\circ$ even if the vacuum mixing angle $\theta_0$ is 
very small. It was pointed out about 12 years ago \cite{ETC,Akh1} that
the probabilities of neutrino flavor transitions can also be strongly
enhanced if the oscillation {\em phase} undergoes certain modification in 
matter. This can happen if the variation of the matter density along the
neutrino path is correlated in a certain way 
with the change of the oscillation phase. This amplification of the neutrino 
oscillation probability in matter due to specific phase relationships has an 
interesting property that it can accumulate if the matter density profile 
along the neutrino path repeats itself, i.e. is periodic. The 
phenomenon is analogous to the resonance in dynamical systems whose parameters 
periodically vary with time -- parametric resonance. It was therefore dubbed 
parametric resonance of neutrino oscillations \cite{ETC,Akh1}. While 
periodicity of the parameters of the system is useful, it is not really
necessary: parametric resonance can occur even in stochastic media (see,
e.g., \cite{stoch}). The stochastic parametric resonance in neutrino
oscillations was briefly discussed in \cite{KS}. 

The parametric resonance can lead to large probabilities of neutrino flavor 
transition in matter even if the mixing angles {\it both} in vacuum {\it and} 
in matter are small. This happens because each half-wave oscillation of the 
transition probability is placed on the top of the previous one, i.e. 
the transition probability builds up (fig. 1). If mixing angle in matter
is very small (matter density is far from the MSW resonance density), the 
parametric resonance enhancement of neutrino oscillations can manifest itself 
only if the neutrinos pass through a large number of periods of density 
modulation, i.e. travel a sufficiently long distance. However, if matter 
density is not very far from the MSW resonance one, an interesting interplay 
between the MSW effect and parametric effects can occur. In particular, a 
strong parametric enhancement of neutrino oscillations can take place even if 
the neutrinos pass only through 1 - 2 periods of density modulation \cite{KS}.

For the parametric resonance to occur, the exact 
shape of the density profile is not very important; what is important is that 
the change in the density be synchronized with the change of the oscillation 
phase. In particular, in \cite{ETC,Akh1} the case of the 
sinusoidal density profile was considered in which the neutrino evolution 
equation reduces to the Mathieu equation. In \cite{Akh1} the parametric 
resonance was also considered for neutrino oscillations in a matter with a 
periodic step function (``castle wall'') density profile, which allows a very 
simple exact analytic solution. We will discuss this solution in sections
2 and 4.  

Although the parametric resonance in neutrino oscillations is certainly an 
interesting physical phenomenon, it requires that very special conditions 
be satisfied. Unfortunately, these conditions cannot be created in the 
laboratory because this would require either too long a baseline or neutrino 
propagation in a matter of too high a density (see sec. 5 below). 
Until recently it was also unclear whether a natural object exists where these
conditions can be satisfied for any known source of neutrinos. This situation 
has changed with a very important observation by Liu and Smirnov \cite{LS} 
(see also \cite{LMS}), who have shown that the parametric resonance conditions 
can be approximately satisfied for the oscillations of atmospheric $\nu_\mu$ 
into sterile neutrinos $\nu_s$ inside the earth. 

It is known that the earth consists of two main structures -- the mantle
and the core. Within the mantle and within the core the matter density changes 
rather slowly (the density variation scale is large compared to the typical
oscillation lengths of atmospheric and solar neutrinos), but at their border it 
jumps sharply by about a factor of two. Therefore to a good approximation one 
can consider the mantle and the core as structures of constant densities equal 
to the corresponding average densities (two-layer model). Neutrinos coming to 
the detector from the lower hemisphere at zenith angles $\Theta$ in the
range defined by $\cos \Theta=(-1) \div (-0.837)$ traverse the earth's mantle, 
core and then again mantle. Therefore such neutrinos experience a periodic
``castle wall'' potential, and their oscillations can be parametrically 
enhanced. Even though the neutrinos pass only through ``1.5 periods'' of 
density modulations (this would be exactly one period and a half if the 
distances neutrinos travel in the mantle and in the core were equal), the 
parametric effects on neutrino oscillations in the earth can be quite
strong. Subsequently, it has been pointed out in \cite{P1} that the parametric 
resonance conditions can also be satisfied (and to even a better accuracy) 
for the $\nu_2\leftrightarrow \nu_{e}$ oscillations in the earth in the case 
of the $\nu_e$ - $\nu_{\mu(\tau)}$ mixing 
\footnote{In \cite{P1} a different name for this phenomenon was suggested --
neutrino oscillation length resonance. We prefer to follow the original 
terminology of \cite{ETC,Akh1}.}. This, in particular, may have important 
implications for the solar neutrino problem. The parametric resonance in the 
oscillations of solar and atmospheric neutrinos in the earth was further 
explored in a number of papers \cite{Akh2,ADLS,CMP,Akh3}. 

In the present paper we review the parametric resonance in neutrino
oscillations and its possible implications for oscillations of solar 
and atmospheric neutrinos in the earth. In sec. 2 we discuss neutrino 
oscillations and their parametric enhancement in matter with ``castle
wall'' density profile. In sec. 3 we discuss the physical interpretation 
of the parametric resonance in neutrino oscillations. 
In sec. 4 the  parametric resonance in oscillations of solar and atmospheric 
neutrinos in the earth is discussed. In sec. 5 the parametric resonance
conditions for neutrino oscillations in the earth are considered. In the
last section the prospects of experimental observation of the parametric 
resonance in neutrino oscillations are discussed and the conclusions are
given.  


\section{Neutrino oscillations in matter with ``castle wall'' density profile} 

Consider oscillations in a 2-flavor neutrino system in a matter with
periodic step function density profile \cite{Akh1,Akh2}. We will be assuming 
that one period of density modulation consists of two parts of the lengths
$T_1$ and $T_2$, with the corresponding effective matter densities $N_1$ and 
$N_2$ (``castle wall'' density profile, figs. 3,5,7). For the $\nu_e
\leftrightarrow \nu_{\mu(\tau)}$ oscillations the effective matter density 
coincides with the electron number density, whereas for the $\nu_{e,\mu,\tau} 
\leftrightarrow \nu_{s}$ oscillations in an isotopically symmetric matter
it is a factor of two smaller. The parametric resonance in such a system
occurs when the oscillations phases $2\phi_1$ and $2\phi_2$ acquired over
the intervals $T_1$ and $T_2$ are odd integer multiples of $\pi$ 
\cite{Akh1,LS,LMS}. Let us denote 
\be
\delta=\frac{\Delta m ^2}{4E}\,,\quad\quad V_i=\frac{G_F}{\sqrt{2}}\,
N_i \,,\quad\quad 
\omega_i=\sqrt{(\cos 2\theta_{0}\,\delta-V_i)^2+
(\sin 2\theta_{0}\,\delta)^2}\,\quad\quad
(i=1,~2)\,.
\label{not}
\ee
Here $E$, $\Delta m^2$ and $\theta_0$ are the neutrino energy, mass squared
difference and vacuum mixing angle, respectively. The difference of the 
neutrino eigenenergies in a matter of density $N_i$ is $2\omega_i$, so that 
the oscillations 
phases acquired over the intervals $T_1$ and $T_2$ are 
\be
2\phi_1 = 2\omega_1 T_1\,,\quad\quad  2\phi_2 = 2\omega_2 T_2\,.  
\label{phases}
\ee
The evolution of a system of oscillating neutrinos is conveniently described 
by the evolution matrix $U$, which in the case of the 2-flavor system is a 
$2\times 2$ unitary matrix. For any interval of time over which the matter
density is constant the evolution matrix can be trivially found; the
evolution matrix in a matter with a step-function density profile is then 
just the product of the corresponding constant-density evolution matrices. 
In particular, for one period of density modulation $T=T_1+T_2$ the
evolution matrix is \cite{Akh2}  
\be
U_T=U_{T_2}U_{T_1}=Y-i\mbox{\boldmath $\sigma$} {\bf X}=
\exp[-i(\mbox{\boldmath $\sigma$}{\bf \hat{X}}) \Phi]\,.
\label{UT2}
\ee
Here $\mbox{\boldmath $\sigma$}$ are the Pauli matrices in the flavor 
space,    
\be
Y=c_1 c_2-({\bf n_1 n_2}) s_1 s_2\,,
\label{Y}
\ee
\be
{\bf X}=s_1 c_2\,{\bf n}_1+s_2 c_1\,{\bf n}_2-s_1 s_2\,
({\bf n}_1\times {\bf n}_2)\,,
\label{X}
\ee
\be
\Phi=\arccos Y=\arcsin X\,,\quad \quad \hat{{\bf X}}=\frac{{\bf X}}{X}\,,
\label{Phi}
\ee 
and we have used the notation
\be
s_i=\sin \phi_i\,,\quad c_i=\cos \phi_i\,,\quad \phi_i=\omega_i T_i\,,
\label{sici}
\ee
\be
{\bf n}_i=
=(\sin 2\theta_i,~0,~-\cos 2\theta_i) \quad \quad \quad (i=1,~2)
\label{ni}
\ee
Here $\theta_i$ is the mixing angle in matter at the density $N_i$. 
Notice that $Y^2+{\bf X}^2=1$ as a consequence of unitarity of $U_T$.

The evolution matrix for $n$ periods ($n$=1, 2,...) can be obtained by
raising $U_T$ to the $n$th power:
\be
U_{nT}
\equiv U(t=nT,~0)
=\exp[-i(\mbox{\boldmath $\sigma$}{\bf\hat{X}}) n\Phi] \,.
\label{UnT}
\ee
Eqs. (\ref{UT2})-(\ref{UnT}) give the exact solution of the evolution 
equation for any instant of time that is an integer multiple of the period
$T$. In order to obtain the solution for $nT<t<(n+1)T$ one has to
evolve the solution at $t=nT$ by applying the evolution matrix
\be 
U_1(t,~nT)=\exp[-iH_1\cdot(t-nT)]
\label{U1t}
\ee
for $nT<t<nT+T_1$ or
\be
U_2(t,~nT+T_1) U_1=\exp[-iH_2\cdot(t-nT-T_1)] \exp[-iH_1 T_1]
\label{U2t}
\ee
for $nT+T_1 \le t < (n+1)T$, with $H_{1,2}$ being the effective 
Hamiltonians of neutrino system at the densities $N_1$ and $M_2$, 
respectively. 

\subsection{Parametric resonance}
Assume that the initial neutrino state at $t=0$ is a flavor eigenstate
$\nu_a$. The probability of finding another flavor eigenstate $\nu_b$ at 
a time $t>0$ (transition probability) is then $P(\nu_a\to \nu_b,~t)=
|U_{21}(t)|^2$ where $U(t)$ is the evolution matrix. For neutrino
oscillations in matter with the ``castle wall'' density profile this
probability can reach its maximum value when the parametric resonance 
conditions are satisfied. These conditions can be written as 
\cite{ETC,Akh1,KS,LS,LMS,P1,Akh2}
\footnote{In refs. \cite{ETC,Akh1,KS} these conditions were derived for 
the particular case $k=k'$, which, however, includes the most important
principal resonance with $k=k'=0$.}
\be
\phi_1=\frac{\pi}{2}+k\pi\,, \quad\quad \phi_2=\frac{\pi}{2}+k'\pi\,,
\quad\quad k, k'=0,1,2,...
\label{rescond}
\ee
At the resonance, the transition probability for the evolution over $n$
periods of density modulation takes a simple form 
\be
P(\nu_a\to \nu_b,~t=nT)=\sin^2 [n(2\theta_2-2\theta_1)] \,.
\label{prob1}
\ee
Let us first assume that the densities $N_1$, $N_2$ are either both below
the MSW resonance density $N_{MSW}$ which is determined from $G_F 
N_{MSW}/\sqrt{2}=\cos 2\theta_0 \,\delta$ or they are both above it. 
This means that the mixing angles $\theta_{1,2}$ satisfy $\theta_{1,2}<\pi/4$ 
or $\theta_{1,2} > \pi/4$, respectively. It is easy to see that in this case 
the difference $2\theta_2-2\theta_1$ is 
always farther away from $\pi/2$ than either $2\theta_1$ or
$2\theta_2$. This means in this case the transition probability
for evolution over one period cannot exceed the maximal transition 
probabilities in matter of constant density equal to either $N_1$ or
$N_2$, namely, $\sin^2 2\theta_1$ or $\sin^2 2\theta_2$. 
However, the parametric resonance does lead to an important gain.
In a medium of constant density $N_i$ the transition probability can never 
exceed $\sin^2 2\theta_i$, no matter how long the distance that neutrinos 
travel. On the contrary, in the matter with ``castle wall'' density
profile, if the parametric resonance conditions (\ref{rescond}) are
satisfied, the transition probability can become large provided neutrinos 
travel large enough distance. 
It can be seen from (\ref{prob1}) that the transition probability can
become quite sizeable even for small $\sin^2 2\theta_1$ and 
$\sin^2 2\theta_2$ provided that neutrinos have traveled sufficiently
large distance. This is illustrated in figs. 2 and 3 for the case  
$N_1,N_2<N_{MSW}$ (the transition probability in the case $N_1,N_2>N_{MSW}$ 
has a similar behavior). The number of periods neutrinos have to pass in 
order to experience a complete (or almost complete) conversion is 
\be
n\simeq \frac{\pi}{4(\theta_1-\theta_2)}\,.
\label{number}
\ee
Consider now the case $N_1<N_{MSW}<N_2$ ($\theta_1<\pi/4<\theta_2$). 
The transition probability over $n$ periods at the parametric resonance 
is again given by eq. (\ref{prob1}). However in this case, for $\theta_2 >
\pi/4 +\theta_1/2$ (which is always satisfied for small mixing in matter), 
one has $\sin^2 (2\theta_2-2\theta_1) > \sin^2 2\theta_1,\, \sin^2
2\theta_2$. This means that {\em even for the time interval equal to one 
period of matter density modulation the transition probability exceeds the 
maximal probabilities of oscillations in matter of constant densities
$N_1$ and $N_2$.} 
This parametric enhancement is further magnified in the case of neutrinos
traveling over ``one and a half'' periods of density modulation, which has 
important implications for neutrinos traversing the earth.
The case $N_1<N_{MSW}<N_2$ is illustrated in figs. 1, 4,5 and 6,7. 

\section{Physical interpretation of the parametric resonance in 
neutrino oscillations }

As we have seen, the parametric resonance can strongly enhance the 
probability of flavor transitions even if the lepton mixing angles both 
in matter and in vacuum is small. This fact can be given a very simple 
physical interpretation 
\footnote{For a geometric interpretation of the parametric resonance in
terms of the spin precession in magnetic field, see \cite{LMS,Sm1,Sm2}. 
The analogy with a pendulum with vertically oscillating point of 
support was discussed in \cite{Akh2}.}. 

Neutrino oscillations in matter of constant density proceed exactly as 
the oscillations in vacuum, the only difference being 
that the oscillation amplitude and length are different from those 
in vacuum. If the vacuum mixing angle $\theta_0$ is small and in addition 
matter density is not close to the MSW resonance one, the amplitude 
of neutrino oscillations in matter, $\sin^2 2\theta$, and therefore the 
transition probability, is small. 

The situation can be drastically different in the case of the ``castle 
wall'' density profile. Consider first neutrino evolution during the 
first part of the period of density modulation (i.e. over the time 
interval $T_1$). The matter density during this interval of time is 
constant: $N(t)=N_1$. If the first of the conditions (\ref{rescond}) is 
satisfied, at the end of this interval the transition probability reaches 
$\sin^2 2\theta_1$ which is the maximal value possible in a matter of
constant density $N_1$. If the density stayed constant, the transition 
probability would have started decreasing and would have returned to zero
at the time $2T_1$. However, at $t=T_1$ the matter density jumps to a value 
$N_2$. If now the second of the conditions in (\ref{rescond}) is also
satisfied, and if in addition 
\be
N_1<N_{MSW}<N_2 \quad\quad (\theta_1 <\pi/4 <\theta_2)\,,
\label{cond1}
\ee
the transition probability will continue increasing instead of starting 
decreasing (figs. 1, 4, 6). In fact, the second half-wave of neutrino
oscillations is similar to the first one. This happens because of the 
violation of the adiabaticity of neutrino oscillations by a sudden change of 
the matter density and because this change is correlated with the change of 
the oscillation phase. The transition probability over one period of
density modulation is 
\be
P(\nu_a\to \nu_b, T)=\sin^2 (2\theta_2-2\theta_1)\,,
\ee
which, as was discussed in the previous section, exceeds both 
$\sin^2 2\theta_1$ and $\sin^2 2\theta_1$ if the condition (\ref{cond1}) is 
satisfied. If the matter density profile is periodic, the increase of the 
transition probability accumulates: The parametric resonance puts each 
half-wave increase of the oscillation curve on the top of the previous
one, leading to a fast growth of the transition probability (fig. 1, 4, 6). 

If the parametric resonance conditions (\ref{rescond}) are satisfied but the 
condition (\ref{cond1}) is not, the transition probability starts decreasing 
after the first half-wave increase. However it does not reach zero, and
the decrease is followed by another increase. As a result, the transition 
probability builds up and can reach unity (fig. 2). In this case, 
however, the increase of the transition probability is less fast than 
when the condition (\ref{cond1}) is satisfied. Similar situation 
takes place if (\ref{cond1}) is obeyed, but the parametric resonance 
conditions (\ref{rescond}) are only approximately satisfied, i.e. there is 
a small detuning. 

We shall now illustrate once again the importance of a correlated change 
of the oscillation phase and matter density profile along the neutrino 
path. In figs. 6 and 7 the coordinate dependence of the transition
probability and matter density profile are shown for a specific case 
in which conditions (\ref{rescond}) and (\ref{cond1}) are fulfilled.
It can be seen from these figures that the probability increase during 
the time intervals $T_2$, which correspond to the effective matter density 
$N_2$, is very small, and, in addition, in this case $T_2 \ll T_1$. 
One could therefore conclude that the evolution during these intervals 
is unimportant. However, this conclusion is wrong: if one removes the 
``spikes'' in the matter density profile of fig. 7, i.e. replaces it by
the profile $N(t)=N_1=const$, the resulting transition probability will be 
very small at all times (fig. 8).

\section{Parametric resonance in neutrino oscillations in the earth}
\subsection{Evolution of oscillating neutrinos in the earth}

As was pointed out in the Introduction, the earth consists of two main 
structures, the mantle and the core, which for the purposes of neutrino 
oscillations can to a very good approximation be considered as layers of
constant density. We shall consider neutrino oscillations in the earth in 
this two-layer approximation. Neutrinos coming to the detector from the 
lower hemisphere of the earth at zenith angles $\Theta$ in the range 
$\cos \Theta=(-1) \div (-0.837)$ (nadir angle $\Theta_n \equiv 180^\circ - 
\Theta \le 33.17^\circ$) traverse the earth's mantle, core and then again 
mantle, i.e. three layers of constant density with the third layer being 
identical to the first one. Therefore such neutrinos experience a periodic 
``castle wall'' potential, and their oscillations can be parametrically 
enhanced. 
Although the neutrinos propagate in this case only through three layers
(``1.5 periods'' of density modulation), the parametric enhancement 
of the transition probability can be very strong. 

The evolution matrix in this case is $U=U_{T_1} U_{T_2} U_{T_1}$. It can
be parametrized as 
\be
U
=Z-i\mbox{\boldmath $\sigma$}{\bf W}\,, \quad\quad Z^2+{\bf W}^2=1\,.
\label{U3}
\ee
The matrix $U$ describes the evolution of an arbitrary initial state and 
therefore contains all the information relevant for neutrino oscillations.  
In particular, the probabilities of the neutrino flavor oscillations $P$ and 
of $\nu_2 \leftrightarrow \nu_{e}$ oscillations $P_{2e}$ are given by 
\cite{Akh2} 
\be
P=W_1^2+W_2^2\,,\quad\quad 
P_{2e}=\sin^2 \theta_0+W_1(W_1 \cos 2\theta_0+W_3 \sin 2\theta_0)\,.
\label{prob}
\ee
We have now to identify the effective densities $N_1$ and $N_2$ with the
average matter densities $N_m$ and $N_c$ in the earth's mantle and core, 
respectively; similarly, we change the notation $V_{1,2}\to V_{m,c}$,
$\phi_{1,2}\to \phi_{m,c}$ and $\theta_{1,2}\to \theta_{m,c}$. 
 
In the two-layer approximation, the parameters $Z$, ${\bf W}$ have a very
simple form \cite{Akh2}: 
\be 
Z=2 \cos\phi_m \,Y-\cos\phi_c \,,
\label{Z}
\ee
\be
{\bf W}=\left(2\sin\phi_m \sin 2\theta_m\,Y+\sin\phi_c\sin 2\theta_c \,,
~0\,,~-\left(2 \sin\phi_m \cos 2\theta_m \,Y + \sin\phi_c \cos 2\theta_c
\right)\right) \,.
\label{W}
\ee
Here the vector ${\bf W}$ was written in components, and and the parameter 
$Y$ was defined in (\ref{Y}). At the parametric resonance, i.e. when the 
conditions (\ref{rescond}) are satisfied, the neutrino flavor transition
probability takes the value \cite{LS,LMS},  
\be
P=\sin^2 (2\theta_c-4\theta_m) \,,
\label{prob2}
\ee
whereas the probability of the $\nu_2 \leftrightarrow \nu_{e}$ transitions is 
\cite{P1} 
\be
P_{2e} = \sin^2 (2\theta_c-4\theta_m+\theta_0) \,.
\label{prob3}
\ee
These probabilities can be close to unity (the arguments of the sines 
close to $\pi/2$) even if the amplitudes of neutrino oscillations in the 
mantle, $\sin^2 2\theta_m$, and in the core, $\sin^2 2\theta_c$, are rather 
small. This can happen if the neutrino energy lies in the range 
$E_c < E < E_m$, where $E_m$ and $E_c$ are the values of the energy that 
correspond to the MSW resonance in the mantle and in the core of the earth.  
This condition is equivalent to the one in eq. (\ref{cond1}). 
The probability $P_{2e}$ is relevant for the description of the oscillations 
of solar neutrinos in the earth \cite{MS2,BW1}. In the case of small mixing
angle MSW solution of the solar neutrino problem, $\sin^2 2\theta_0 < 10^{-2}$ 
\cite{BKS}, and $P_{2e}$ practically coincides with $P$ unless both 
probabilities are very small.  

The trajectories of neutrinos traversing the earth are determined by their 
nadir angle $\Theta_n=180^\circ-\Theta$. The distances $R_m$ and $R_c$
that neutrinos travel in the mantle (each layer) and in the core are given by  
\be
R_m=R\left(\cos\Theta_n-\sqrt{r^2/R^2-\sin^2\Theta_n}\,\right)\,,
\quad\quad R_c=2R \,\sqrt{r^2/R^2-\sin^2\Theta_n}\,.
\label{RmRc}
\ee
Here $R=6371$ km is the earth's radius and $r=3486$ km is the radius of the 
core. The matter density in the mantle of the earth ranges from 2.7 $g/cm^3$ 
at the surface to 5.5 $g/cm^3$ at the bottom, and that in the core ranges
from 9.9 to 12.5 $g/cm^3$ (see, e.g., \cite{Stacey}). The electron number 
fraction $Y_e$ is close to 1/2 both in the mantle and in the core.
Taking the average matter densities in the mantle and core to be 4.5 and 11.5
$g/cm^2$ respectively, one finds for the $\nu_e\leftrightarrow \nu_{\mu,\tau}$
oscillations involving only active neutrinos the following values of $V_m$
and $V_c$: $V_m=8.58\times 10^{-14}$ eV, $V_c=2.19\times 10^{-13}$ eV.
For transitions involving sterile neutrinos $\nu_e\leftrightarrow
\nu_{s}$ and $\nu_{\mu,\tau}\leftrightarrow \nu_s$, these parameters
are a factor of two smaller.

\subsection{Parametric resonance conditions for neutrino oscillations in
the earth}

If the parametric resonance conditions (\ref{rescond}) are satisfied, strong 
parametric enhancement of the oscillations of core crossing neutrinos in the 
earth can occur \cite{LS,LMS,P1,Akh2,ADLS,CMP}, see fig. 1  
\footnote{For the parametric resonance to be a maximum of the transition
probability, an additional condition has to be satisfied 
(two conditions in the case of $\nu_2 \leftrightarrow \nu_{e}$ oscillations
\cite{P1}). For the oscillations between neutrinos of different flavor, this 
condition can be written as $\cos(2\theta_c-4\theta_m)<0$. For small vacuum 
mixing angles $\theta_0$ it essentially reduces to $\delta <V_c$, with a 
small region around $\delta=V_m$ excluded.}. 
We shall now discuss these conditions. The phases $\phi_m$ and $\phi_c$ depend 
on the neutrino parameters $\Delta m^2$, $\theta_0$ and $E$ and also on the 
distances $R_m$ and $R_c$ that the neutrinos travel in the mantle and in the 
core. The path lengths $R_m$ and $R_c$ vary with the nadir angle; however,
as can be seen from (\ref{RmRc}), their changes are correlated and they cannot 
take arbitrary values. Therefore if for some values of the neutrino parameters 
a value of the nadir angle $\Theta_n$ exists for which, for example, the first
condition in eq. (\ref{rescond}) is satisfied, it is not obvious if at the same 
value of $\Theta_n$ the second condition will be satisfied as well. In other 
words, it is not clear if the parametric resonance conditions can be fulfilled 
for neutrino oscillations in the earth for at least one set of the neutrino 
parameters $\Delta m^2$, $\theta_0$ and $E$. However, as was shown in 
\cite{P1,Akh2}, not only the parametric resonance conditions are satisfied (or 
approximately satisfied) for a rather wide range of the nadir angles covering 
the earth's core, they are fulfilled for the ranges of neutrino parameters 
which are of interest for the neutrino oscillations solutions of the solar and
atmospheric neutrino problems. In particular, the conditions for the principal 
resonance ($k=k'=0$) are satisfied to a good accuracy for $\sin^2 2\theta_0 
\aprle 0.1$, $\delta \simeq (1.1 \div 1.9)\times 10^{-13}$ eV$^2$, which 
includes the ranges relevant for the small mixing angle MSW solution of the 
solar neutrino problem and for the subdominant $\nu_\mu \leftrightarrow 
\nu_{e}$ and $\nu_e \leftrightarrow \nu_{\tau}$ oscillations of atmospheric 
neutrinos 
\footnote{The parametric enhancement of neutrino oscillations in the earth 
for neutrinos that travel significant distances in the earth's core was
found numerically in a number of earlier papers \cite{numer}. However, in
these papers the parametric nature of the effect has not been recognized. }. 

The fact that the parametric resonance conditions can be satisfied so well for 
neutrino oscillations in the earth is rather surprising. It is a consequence 
of a number of remarkable numerical coincidences. It has been known for some 
time \cite{GKR,LS,LiLu} that the potentials $V_m$ and $V_c$   
corresponding to the matter densities in the mantle and core, the inverse
radius of the earth $R^{-1}$, and typical values of $\delta\equiv
\Delta m^2/4E$ of interest for solar and atmospheric neutrinos, are all of
the same order of magnitude -- ($3\times 10^{-14}$ -- $3\times 10^{-13}$) eV.
It is this surprising coincidence that makes appreciable earth effects on the 
oscillations of solar and atmospheric neutrinos possible. However, for the
parametric resonance to take place, a coincidence by an order of magnitude is 
not sufficient: the conditions (\ref{rescond}) have to be satisfied at least 
within a 50\% accuracy \cite{Akh2}. This is exactly what takes place. In 
addition, in a wide range of the nadir angles $\Theta_n$, with changing 
$\Theta_n$ the value of the parameter $\delta$ at which the resonance 
conditions (\ref{rescond}) are satisfied slightly changes, but the fulfillment 
of these conditions is not destroyed. 

In this row of mysterious coincidences, at least the last one -- the stability 
of the parametric resonance conditions with respect to variations of the nadir 
angle -- has a simple explanation. It is related to the fact that, due to 
the spherical geometry of the earth, with increasing nadir angle $R_m$ 
increases and $R_c$ decreases so that in a large interval of the nadir angles 
covering the earth's core that the sum $1/R_c +1/R_m$ is almost constant. 
For more details, see ref. \cite{Akh3}. 

The parametric enhancement of neutrino oscillations in the earth can also 
occur when either $k$ or $k'$ in eq. (\ref{rescond}) or both are different 
from zero (higher-order parametric resonances). However, the corresponding 
resonance conditions can only be satisfied for the values of neutrino 
mass squared differences and mixing angles which are of no practical 
interest for any known source of neutrinos, possible exception being the
$hep$ component of the solar neutrino flux \cite{Akh3}. 

\section{Can the parametric resonance in neutrino oscillations be
observed?}
Besides being an interesting physical phenomenon, the parametric resonance in  
neutrino oscillations can provide us with an important additional information  
about neutrino properties. 
Therefore experimental observation of this effect would be of considerable 
interest. We shall now discuss the prospects for experimental observation 
of the parametric resonance in neutrino oscillations in the earth, having 
in mind mainly the principal resonance. There are two main sources of neutrinos 
for which the parametric resonance can be important -- atmospheric neutrinos 
and solar neutrinos. Both sources have their advantages and disadvantages 
from the point of view of the possibility of observation of the parametric 
resonance. We shall now briefly discuss them. 

We start with atmospheric neutrinos. The parametric resonance can occur in the 
$\nu_\mu \leftrightarrow \nu_s$ \cite{LS,LMS} and also in the subdominant 
$\nu_e \leftrightarrow \nu_{\mu(\tau)}$ channels of oscillations \cite{ADLS}.  
It can affect the distributions of $\mu$-like events and also (in the case of 
the $\nu_e \leftrightarrow \nu_{\mu(\tau)}$ oscillations) lead to interesting 
peculiarities in the zenith angle distributions of the multi-GeV e-like events. 

The observation of the parametric effects is hampered 
by the loose correlation between the directions of the momenta of atmospheric
neutrinos and of the charged leptons which they produce and which are actually 
detected. Because of this the trajectories of neutrinos coming to the detector 
are not known very precisely. In addition, the data are presented for certain 
samples of events (sub-GeV, multi-GeV, upward through-going, upward stopping) 
which includes collecting data over rather wide energy intervals. The 
contributions of the parametric peaks may therefore be integrated over together 
with other possible enhancement peaks -- due to the MSW resonances in the 
mantle and in the core, making the distinction between these effects 
difficult. Also, strong resonance enhancement effects (both parametric and 
MSW) can only occur either for neutrinos or for antineutrinos, depending on 
the sign of $\Delta m^2$ 
\footnote{In general, the parametric enhancement can take place even for 
``wrong sign'' $\Delta m^2$, but in the case of neutrino oscillations in the 
earth these effects are small.}. 
The present atmospheric neutrino experiments do not distinguish between 
neutrinos and antineutrinos, therefore possible matter effects are ``diluted'' 
in the sum of the $\nu$- and $\bar{\nu}$-induced events. At certain values of 
the ratio of the muon and electron neutrino fluxes $r(E,\Theta_n)$ depending on 
the value of mixing angle $\theta_{23}$ the parametric effects on e-like events 
are suppressed \cite{ADLS}. 

Atmospheric neutrinos have some advantages for observation of the parametric 
resonance in neutrino oscillations in the earth. Neutrinos come to the
detectors from all directions, which means that practically the whole solid 
angle covering the earth's core will contribute to the effect. There are no 
additional suppression factors due to a specific composition of the incoming  
neutrino flux which may quench the earth's matter effect on the oscillations 
of solar neutrinos (see below). Parametric effects may provide a sensitive  
probe of the neutrino mixing angle $\theta_{13}$ with sensitivity possibly 
going beyond that of the long-baseline accelerator and reactor experiments 
\cite{ADLS,ALLS}. Possible ways of improving the prospects of the experimental 
observation of the parametric effects in the atmospheric neutrino oscillations 
include using various energy cuts, finer zenith angle binning and detectors 
capable of detecting the recoil nucleon, which would enable one to reconstruct 
the direction of an incoming neutrino \cite{ADLS,ALLS}. It would also be 
highly desirable to have detectors that can determine the charge of the
observed electrons and muons, i.e. discriminate between neutrinos and 
antineutrinos. 

Solar neutrinos can experience a strong parametric enhancement of their 
oscillations in the earth if the small mixing angle MSW effect is the correct 
explanation of the the solar neutrino deficit \cite{P1,Akh2}. The parametric 
enhancement can occur in a wide range of values of $\sin^2 2\theta_0$ and for 
the nadir angles $\Theta_n$ almost completely covering the core of the earth. 
The trajectory of each detected neutrino is exactly known. For boron neutrinos 
the resonance occurs at the values of $\Delta m^2$ which correspond to the 
central part of the allowed interval for the small mixing angle MSW solution 
of the solar neutrino problem. 

However, there are some disadvantages, too. 
Unfortunately, due to their geographical  location, the existing solar
neutrino detectors have a relatively low time during which
solar neutrinos pass through the core of the earth to reach the detector
every calendar year. The Super-Kamiokande detector has a largest fractional
core coverage time equal to $7\%$. In \cite{GKR} it was suggested to build
a new detector close to the equator in order to increase the sensitivity   
to the earth regeneration effect; this would also maximize the parametric
resonance effects in oscillations of solar neutrinos in the earth.
In the case of the MSW solutions of the solar neutrino problem the probability 
$P_{SE}$ of finding a solar $\nu_e$ after it traverses the earth depends 
sensitively on the average $\nu_e$ survival probability in the sun $\bar{P}_S$ 
\cite{MS2,BW1}:
\be
P_{SE}=\bar{P}_S + \frac{1-2 \bar{P}_S}{\cos 2\theta_0}\,(P_{2e}-\sin^2 \theta_0)\,.
\label{PSE}
\ee 
The probability $P_{2e}$ can experience a strong parametric enhancement, but 
in the case of small mixing angle MSW solution of the solar neutrino problem  
the probability $\bar{P}_S$ for the Super-Kamiokande and SNO experiments 
turns out to be rather close to 1/2. This means that the effects of passage 
through the earth on solar neutrinos should be strongly suppressed. The 
current best fit of the solar neutrino data is not far from the line  in the 
parameter space where $\bar{P}_S$ is exactly equal to 1/2 and $P_{SE}=
\bar{P}_S$ (i.e. the earth matter effects are absent). Whether or not it will 
be possible to observe the parametric resonance in the oscillations of solar 
neutrinos in the earth depends on how close to this line the true values of 
$\sin^2 2\theta_0$ and $\Delta m^2$ are. By now the Super-Kamiokande 
experiment has not observed, within its experimental accuracy, any enhancement 
of neutrino signal for earth core crossing neutrinos \cite{SK2}. This can be 
because the parametric enhancement of the neutrino oscillations in the 
earth does not occur (e.g.  if the true solution of the solar neutrino 
problem is vacuum oscillations or large mixing angle MSW effect), or because 
the values of $\Delta m^2$ and $\sin^2 2\theta_0$ are too close to those at 
which $\bar{P}_S=1/2$. Hopefully, with accumulated statistics of the 
Super-Kamiokande and forthcoming data from the SNO experiment the situation 
will soon be clarified. 

It is interesting to note that the Super-Kamiokande data on the zenith angle 
dependence of the solar neutrino events seems to indicate some deficiency 
of the events due to the core-crossing neutrinos rather than an excess  
\cite{SK2}, although it is not statistically significant. Should this 
deficiency be confirmed by future data with better statistics, it could have 
a natural explanation in terms of the parametric resonance of neutrino 
oscillations. As follows from (\ref{PSE}), the parametric enhancement of 
$P_{2e}$ for core crossing neutrinos can lead to a deficiency of the events 
if the neutrino parameters are in the small-$\sin^2 2\theta_0$ part of the 
allowed region which corresponds to $P_S>1/2$ (see, e.g., fig. 10 in 
ref. \cite{BK}). In this case one should also have an ``opposite sign'' overall 
day-night effect (fewer events during the night than during the day). 
In any case, given the current experimental constraints 
on the neutrino parameters, if the small mixing angle MSW effect is the true 
solution of the solar neutrino problem, the only hope to observe earth matter 
(day-night) effects on solar neutrinos seems to be through the parametric 
resonance of oscillations of core crossing neutrinos. 

As we have seen, observing the parametric resonance in oscillations of solar 
and atmospheric neutrinos in the earth is not an easy task. Can one create 
the necessary matter density profile and observe the parametric resonance in 
neutrino oscillations in the laboratory (i.e. short-baseline) experiments?  
Unfortunately, the answer to this question seems to be negative. The parametric 
resonance can occur when the mean oscillation length in matter approximately 
coincides with the matter density modulation length \cite{ETC,Akh1,KS}: 
$l_m\simeq L$. In a matter of density $N_i$ the oscillation length is given by
$l_m=\pi/\omega_i$ where $\omega_i$ was defined in (\ref{not}). Let us require 
$l_m \aprle 1$ km. Assume first that $V_i\aprge\delta$, i.e. $\omega_i$ are 
dominated by matter density terms. Than for $l_m\aprle 1$ km one would need 
a matter of mass density $\rho_i\ge 3.3\times 10^4$ g/cm$^3$, clearly not a 
feasible value. Conversely, for $\rho_i\le 10$ g/cm$^3$, one finds $l_m\aprge 
3300$ km, a distance comparable with the earth's radius. Consider now the 
opposite case, $\delta \gg V_i$. Then the oscillation length in matter 
essentially coincides with the vacuum oscillations length which in principle 
can be rather short provided that the vacuum mixing angle $\theta_0$ is small 
(otherwise this would contradict reactor and accelerator data). However, in 
this case there is another problem. Requiring $l_m\aprle 1$ km one finds 
$\delta \aprge 2.5 \times 10^{-10}$ eV. For $\rho_i\aprle 10$ g/cm$^3$ one 
therefore has $V_i/\delta \aprle 10^{-3}$. This means that the mixing angles 
in matter are very close to the vacuum one, $\theta_i\simeq \theta_0 
(1+V_i/\delta)$, and so their difference is very small: $\Delta \theta =
\theta_2-\theta_1 \simeq (\Delta V/\delta)\,\theta_0\aprle 10^{-3}\theta_0$.  
When the difference of mixing angles in matter is small, the parametric 
effects can manifest themselves only if neutrinos travel over a large 
number of periods, $n\simeq \pi/4 \Delta\theta$ [see (\ref{number})]. 
Therefore in this case the necessary baseline is $\sim \pi^2/(4\theta_0 
\Delta V) \aprge 3\times 10^3$ km, again too large. One can conclude that 
the sole presently known object where the parametric resonance in neutrino 
oscillations can take place is the earth, as was first pointed out in
\cite{LS,LMS}. 


This work was supported by Funda\c{c}\~ao para a Ci\^encia e a Tecnologia 
through the grant PRAXIS XXI/BCC/16414/98 and also in part by the TMR network 
grant ERBFMRX-CT960090 of the European Union.

\newpage
\centerline{\large Figure captions}

\vglue 0.4cm
\noindent
Fig. 1. 
Solid curve: transition probability $P$ for 
$\nu_e\leftrightarrow \nu_{\mu,\tau}$ oscillations 
in the earth as a function of the distance $t$ (measured in units of
the earth's radius $R$) along the neutrino trajectory. 
$\delta\equiv \Delta m^2/4E =1.8\times 
10^{-13}$ eV, $\sin^2 2\theta_{0}=0.01$, $\Theta_n=11.5^\circ$. Dashed 
curve: the same for a hypothetical case of neutrino propagation over 
full two periods of density modulation ($t_{max}=2(R_m+R_c$)).

\noindent
Fig. 2. 
Coordinate dependence of the neutrino flavor transition probability 
$P$ in a matter with the castle wall density profile. 
$\sin^2 2\theta_{0}=0.01$, 
$\delta= 10^{-12}$ eV, 
$V_1=10^{-13}$ eV, $V_2=6.33\times 10^{-13}$ eV, 
$T_1=5.4\times 10^{-2}$, $T_2=0.1296$, 
all distances are in units of $R=3.23 \times 10^{13}$ eV$^{-1}$. 

\noindent
Fig. 3. 
Coordinate dependence of the matter-induced neutrino potential 
[$(G_F/\sqrt{2}\times$ (density profile)] for the case shown in fig. 2.  

\noindent
Fig. 4. 
Same as fig. 2 but for $\delta=2\times 10^{-13}$ eV, 
$V_2=4\times 10^{-13}$ eV, 
$T_1=0.4814$, $T_2=0.2408$,

\noindent
Fig. 5. 
Coordinate dependence of the matter-induced neutrino potential 
for the case shown in fig. 4.  

\noindent
Fig. 6. 
Same as fig. 2 but for $\delta=10^{-12}$ eV, 
$V_2=10^{-11}$ eV, $T_1=5.4\times 10^{-2}$, $T_2=5.4\times 10^{-3}$,

\noindent
Fig. 7. 
Coordinate dependence of the matter-induced neutrino potential 
for the case shown in fig. 6.

\noindent
Fig. 8. 
Same as fig. 6 but for $V_2=V_1$ ($V(t)=V_1=const$).

\newpage
\begin{figure}[H]
\hglue -2.0cm
\mbox{\epsfig{figure=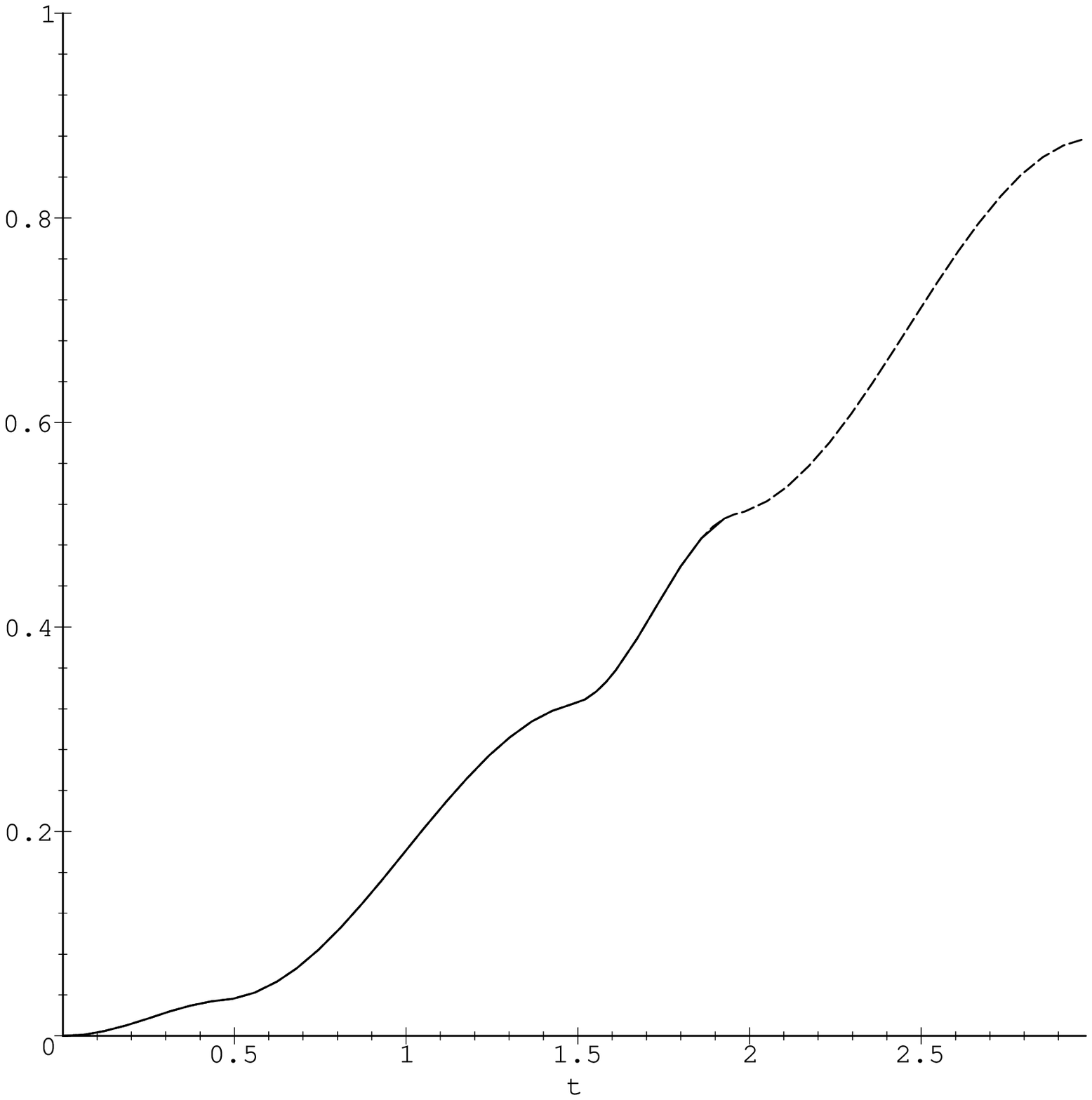,width=19.0cm, height=19.0cm}}
\vglue -1.5cm
\centerline{\mbox{Fig. 1.}}
\end{figure}

\newpage
\begin{figure}[H]
\vglue -1cm 
\mbox{\epsfig{figure=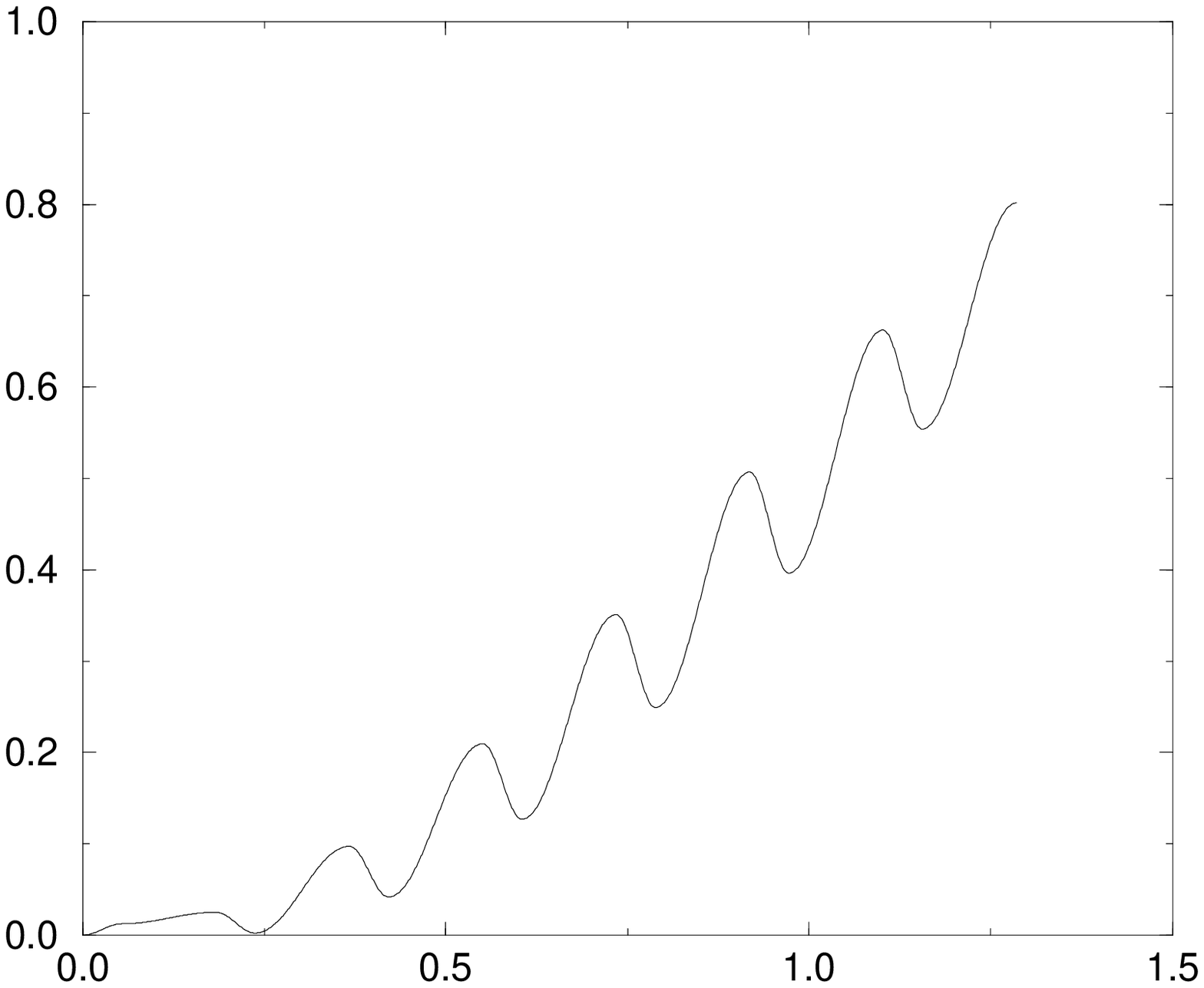,width=15.0cm, height=12.0cm}}
\vglue -1.5cm
\centerline{\mbox{Fig. 2.}}
\mbox{\epsfig{figure=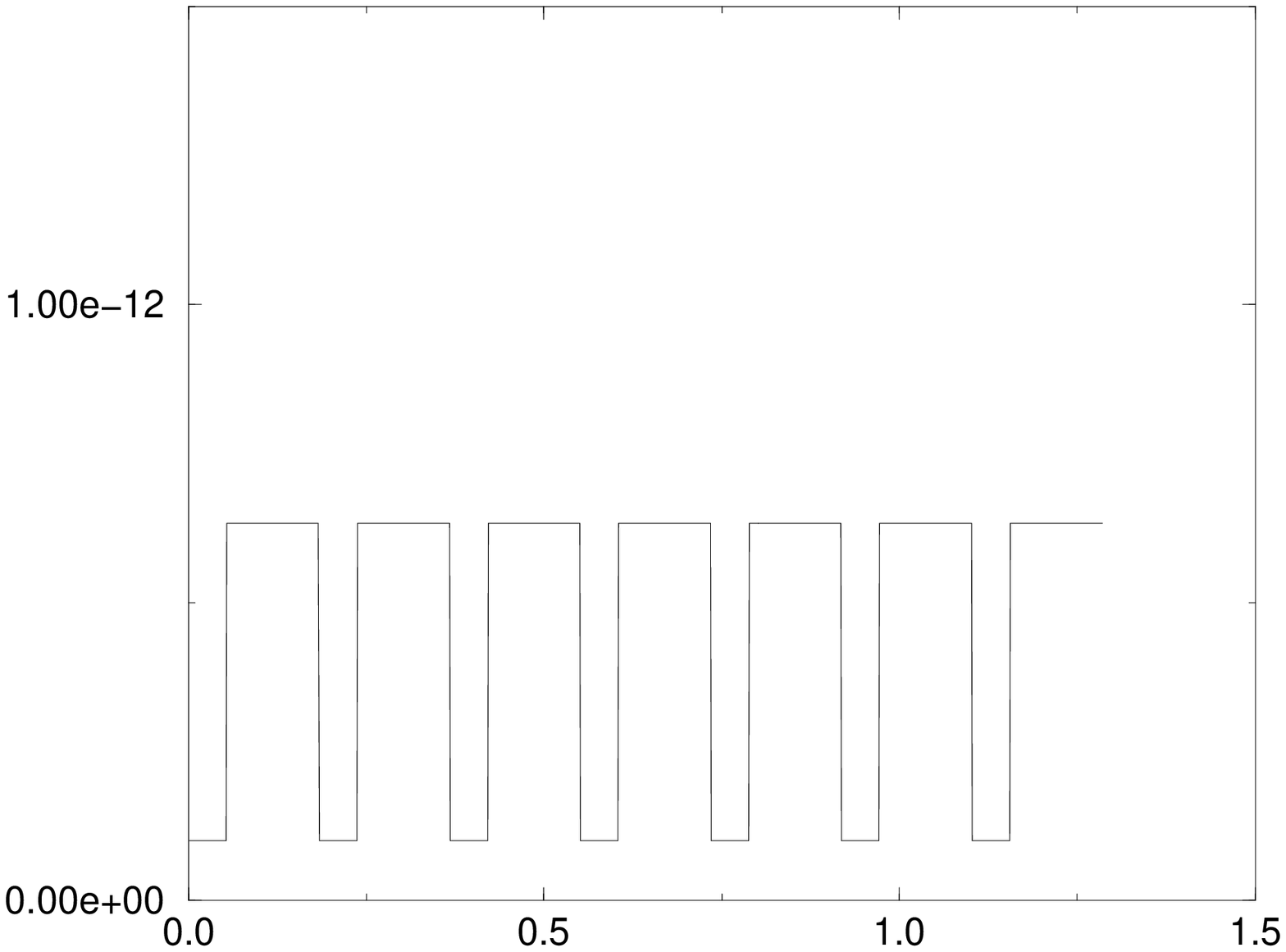,width=15.0cm, height=12.0cm}}
\vglue -1.5cm
\centerline{\mbox{Fig. 3.}}
\end{figure}

\newpage
\begin{figure}[H]
\vglue -1cm 
\mbox{\epsfig{figure=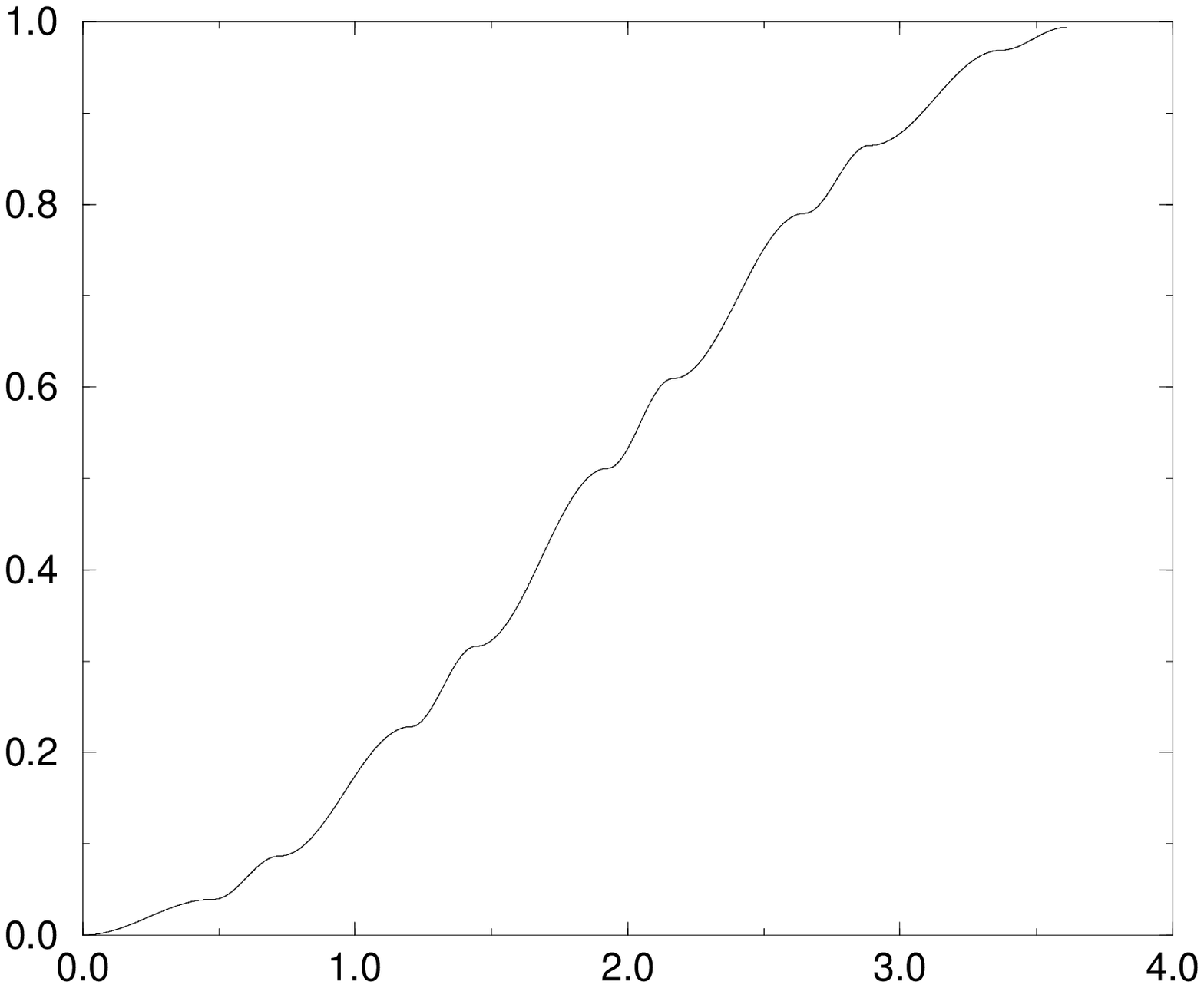,width=15.0cm, height=12.0cm}}
\vglue -1.5cm
\centerline{\mbox{Fig. 4.}}
\mbox{\epsfig{figure=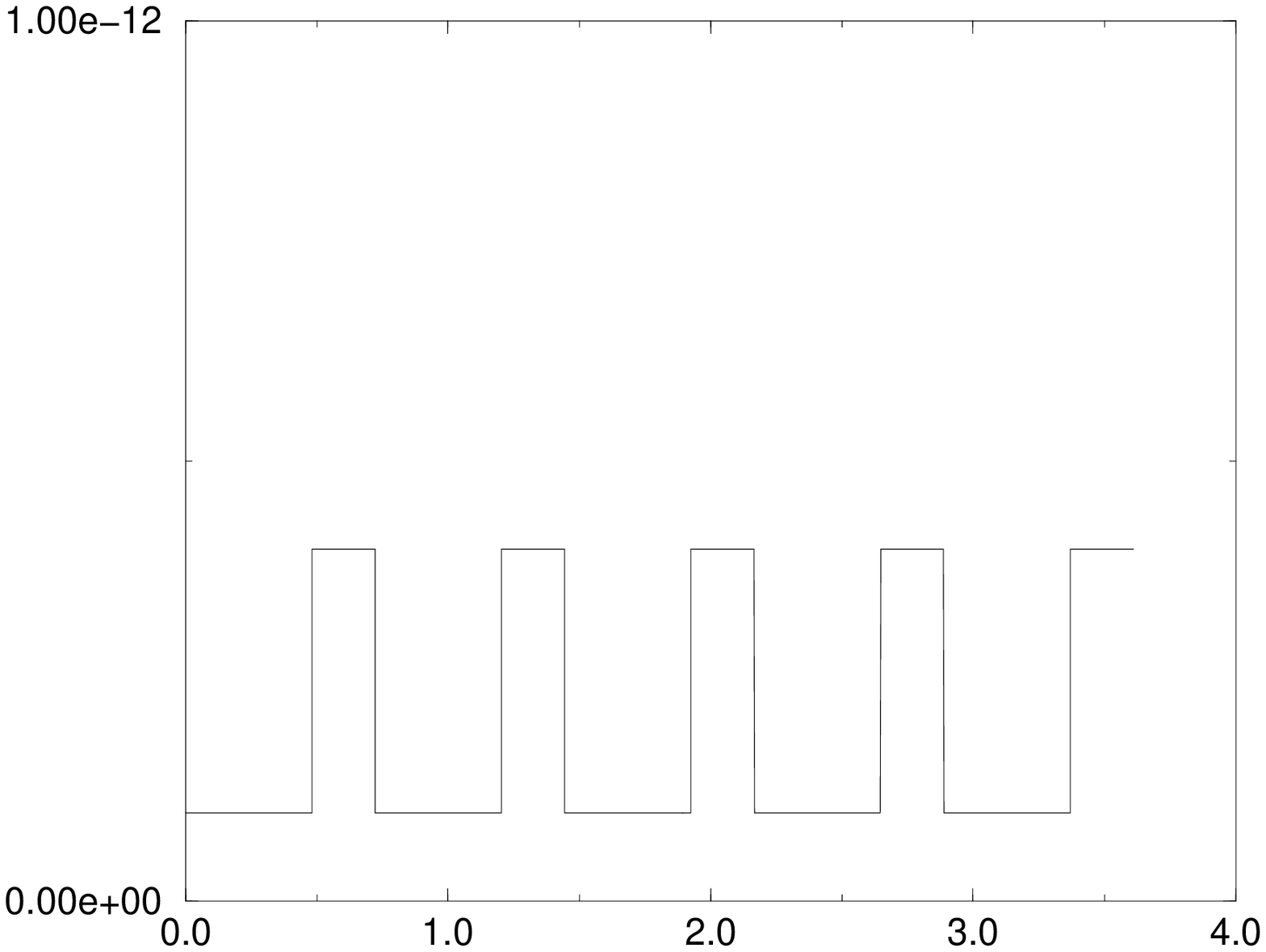,width=15.0cm, height=12.0cm}}
\vglue -1.5cm
\centerline{\mbox{Fig. 5.}}
\end{figure}

\newpage
\begin{figure}[H]
\vglue -1cm 
\mbox{\epsfig{figure=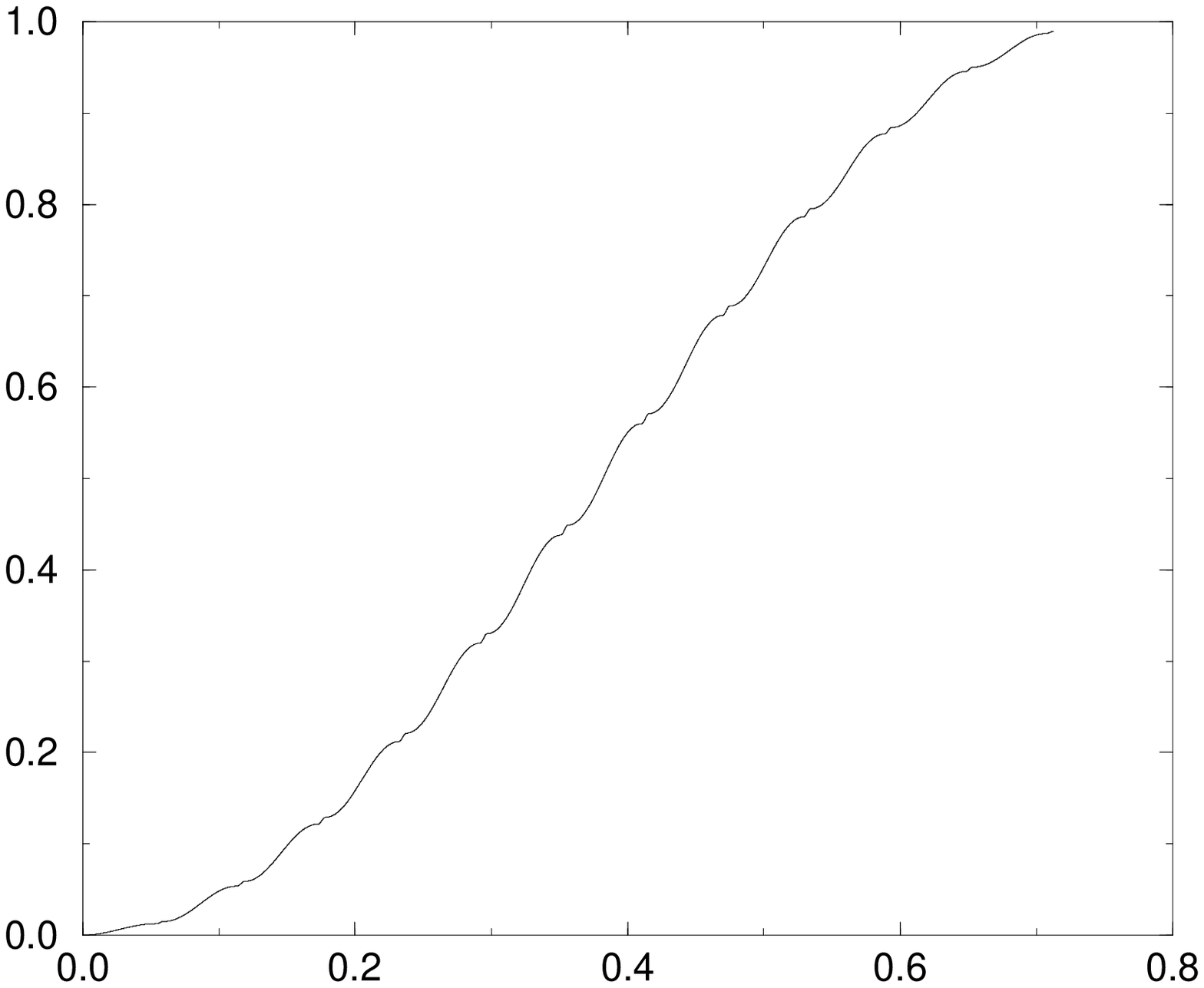,width=15.0cm, height=12.0cm}}
\vglue -1.5cm
\centerline{\mbox{Fig. 6.}}
\mbox{\epsfig{figure=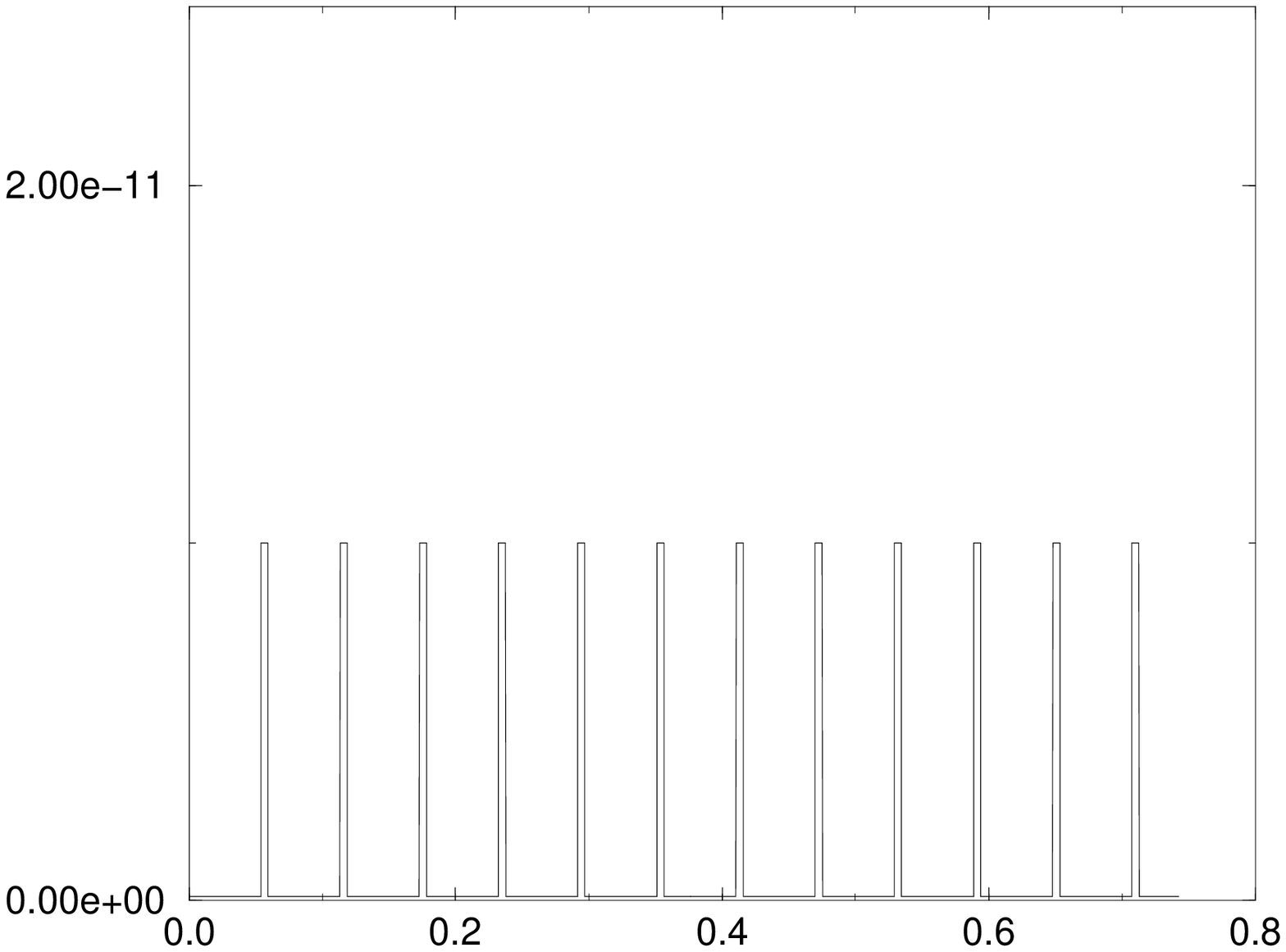,width=15.0cm, height=12.0cm}}
\vglue -1.5cm
\centerline{\mbox{Fig. 7.}}
\end{figure}

\newpage
\begin{figure}[H]
\vglue -1cm 
\mbox{\epsfig{figure=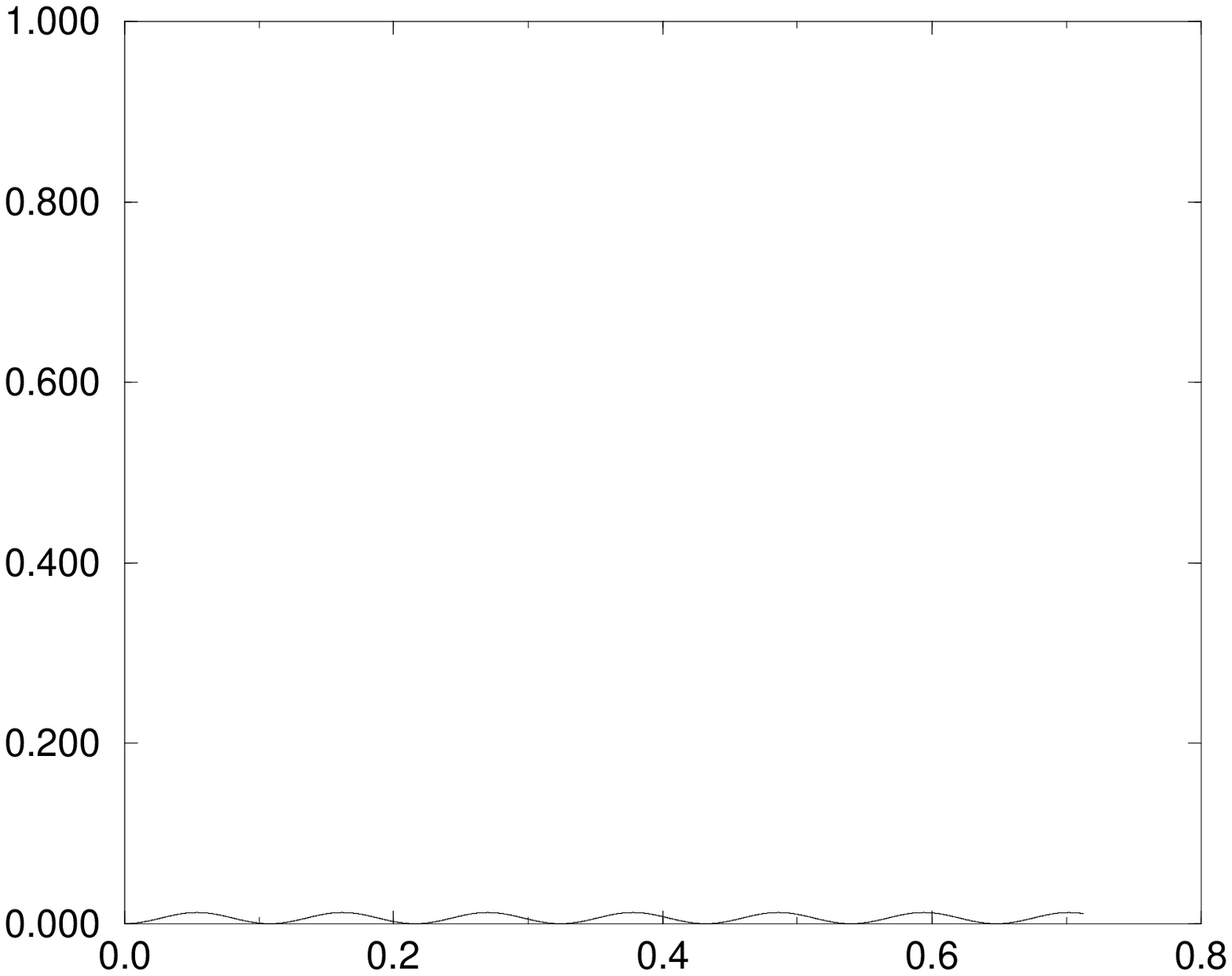,width=15.0cm, height=12.0cm}}
\vglue -1.5cm
\centerline{\mbox{Fig. 8.}}
\vglue -1.5cm
\end{figure}

\end{document}